\documentclass[12pt]{article}

\usepackage[T1]{fontenc}
\usepackage{graphicx}
\usepackage{mathrsfs}
\usepackage{amssymb}
\usepackage{mathrsfs}
\usepackage{amssymb}
\usepackage{txfonts}

\newfont{\mbld}{cmbx10 scaled 800}
\newfont{\cab}{cmsy10 scaled 1200}
\newfont{\scab}{cmsy10 scaled 1000}
\newfont{\bcall}{cmbsy10 scaled 1200}

\def\XXint#1#2#3{{\setbox0=\hbox{$#1{#2#3}{\int}$ }
\vcenter{\hbox{$#2#3$ }}\kern-.65\wd0}}

\begin{document}
\title{\Large\bf On a path integral representation of the 
\\[3pt]
Nekrasov instanton partition function  
\\[3pt]
and its Nekrasov--Shatashvili limit}
\author{\normalsize
Franco Ferrari$^1$\footnote{e-mail:
    ferrari@fermi.fiz.univ.szczecin.pl}
$\;$ and $\;$ Marcin Pi\c{a}tek$^{1,2}$\footnote{e-mail:
  piatek@fermi.fiz.univ.szczecin.pl} \\
\normalsize  $^1$ Institute of Physics and CASA*, University of Szczecin,\\
\normalsize Wielkopolska 15, 70451 Szczecin, Poland\\
\normalsize  $^2$ Bogoliubov Laboratory of Theoretical Physics,\\
\normalsize Joint Institute for Nuclear Research, 141980, Dubna, Russia
}
\maketitle
\abstract{
In this work we study the Nekrasov--Shatashvili limit of the Nekrasov
instanton partition function of  Yang--Mills
field theories with ${\cal N}=2$ supersymmetry and gauge group
$SU(N)$. The theories are coupled with fundamental matter.
A path integral expression of the full instanton partition function
is derived. It is checked that in the Nekrasov--Shatashvili (thermodynamic)
limit the action of the field theory obtained in this way reproduces
exactly the  equation of
motion used in the saddle-point calculations.
}

\section{Introduction}
In this letter we derive a path integral expression of the
Nekrasov multi--instanton partition function
$Z_{\sf inst}(q,\epsilon_1,\epsilon_2)$ that takes into account the
instantonic sector of gauge field theories with ${\cal N}=2$
supersymmetry and gauge group $SU(N)$.
Here $q$ denotes an effective scale, while $\epsilon_1$ and
$\epsilon_2$ are the
deformation parameters of the $\Omega-$background in which the ${\cal
  N}=2$ gauge theory has been embedded before applying the
localization procedure.
More details on the derivation of
$Z_{\sf inst}(q,\epsilon_1,\epsilon_2)$ can be found in
Refs.~\cite{Nekrasov1,NekrasovOkounkov,NekraShadchin,Marshakov1}.
The Nekrasov multi--instanton partition function plays a crucial role
in the so--called AGT-W  \cite{AGT,Wyllard,MM}
and Bethe/gauge correspondences \cite{NekraSha,NekraSha2,NekraSha3}.

To provide a  representation of $Z_{\sf
  inst}(q,\epsilon_1,\epsilon_2)$ in the form
of the partition function of scalar fields we apply techniques borrowed
from the theory of matrix models \cite{Parisi}.
Usually the multi--instanton partition function is given in the form
of a sum over $k$ variables $\phi_I$, $I=1,\ldots,k$, called eigenvalues,
where the integer $k$
ranges from zero to infinity. In the path integral formulation,
one scalar field is related to the density of eigenvalues and the
second is a lagrange multiplier.
Let us note that matrix models methods have already been applied in the past
to investigate the instantonic partition function in different
contexts, see for instance
\cite{Dijkgraaf,Sulkowski,IO,MMS,MMM,MMM2,MMS3,MMS4,Morozov}, 
but up to now not to the problem
of deriving a path integral expression of
$Z_{\sf inst}(q,\epsilon_1, \epsilon_2)$.
Starting from this expression we are able to prove in a simple and
rigorous way that in the Nekrasov--Shatashvili limit $\epsilon_2\to 0$,
performed by keeping $\epsilon_1$ constant,
the full multi--instanton
partition function reduces to its saddle point approximation given in
\cite{Fucitoiinni}.
This limit is relevant in a wide set of applications, see for example
\cite{NekraSha,Fucitoiinni,Pogho1,BMT,Dorey,Dorey2,Piatek1,FFPiatek}.
With the series representation of $Z_{\sf inst}(q,\epsilon_1,
\epsilon_2)$ the computation of the Nekrasov--Shatashvili limit is
particularly cumbersome, because it requires
to extract from $Z_{\sf inst}(q,\epsilon_1, \epsilon_2)$
the contribution
coming from
the tree diagrams, which leads to rather complicated calculations.
Within the path integral representation
the limiting procedure becomes straightforward since it is performed
directly inside the action of the theory.

\section{Path integral formulation}
The starting point is the instanton partition function of a ${\cal
  N}=2$ gauge field theory with gauge group $SU(N)$ and matter in the
fundamental representation \cite{NekrasovOkounkov,Fucitoiinni}:
\begin{eqnarray}
Z_{\sf inst}(q,\epsilon_1,\epsilon_2)&=&
1+\sum_{k=1}^\infty \frac{q^k}{k!}\left(\frac{\epsilon_1+\epsilon_2}{
  \epsilon_1\epsilon_2}\right)^k\int\limits_{\mathbb{R}}\prod_{I=1}^k\frac{d\phi_I}{2\pi
i}\prod_{I\ne J=1}^kD(\phi_I-\phi_J)\nonumber\\
&\times&\prod_{I=1}^kQ(\phi_I),
\label{zinst}
\end{eqnarray}
where
\begin{equation}
D(z)=\frac{z(z+\epsilon_1+\epsilon_2)}{(z+\epsilon_1)(z+\epsilon_2)},\qquad\qquad
Q(z)=\frac{M(z)}{P(z+\epsilon_1+\epsilon_2)P(z)}
\end{equation}
and
\begin{equation}
M(z)=\prod_{r=1}^{N_f}(z+m_r),
\qquad\qquad
P(z)=\prod_{l=1}^N(z-a_l).
\end{equation}
Here the $a_l$'s, $l=1,\ldots,N$, are the vacuum expectation values of
the adjoint scalar field in the $SU(N)$ vector multiplet, while the
$m_r$'s parametrize the masses of the fundamental matter.
$N_f$ can be indentified with the number of flavors of the theory,
though a more precise description of its meaning can be found in
\cite{Nekrasov1}. Let us note that the integrals over the real line
$\mathbb{R}$ in Eq.~(\ref{zinst}) require some form of regularization
and should be intended as integrals over a closed contour. The details
are explained in \cite{Nekrasov1} and, for generic matrix models, in
\cite{Parisi}.
 $Z_{\sf inst}(q,\epsilon_1,\epsilon_2)$
can be rewritten in a more appropriate way
for our purposes:
\begin{equation}
Z_{\sf inst}(q,\epsilon_1,\epsilon_2)=\sum_{k=0}^{\infty}
\frac{q^k}{k!}
\left(\frac{\epsilon_1+\epsilon_2}{\epsilon_1\epsilon_2}\right)^k
\int\limits_{\mathbb{R}} \prod_{I=1}^k\frac{d\phi_I}{2\pi i}
\;{\rm e}^{H_k}, \label{zinsrew}
\end{equation}
where
\begin{equation}
H_k=\sum_{I\ne
J=1}^k\log(D(\phi_I-\phi_J))+\sum_{I=1}^k\log(Q(\phi_I)).
\end{equation}
After introducing the density of eigenvalues
\begin{equation}
\rho(\phi)=\sum_{I=1}^k
\frac{\epsilon_1\epsilon_2}{\epsilon_1+\epsilon_2}
\,\delta(\phi-\phi_I)
\label{eigdensdef}
\end{equation}
it is possible to express $Z_{\sf inst}(q,\epsilon_1,\epsilon_2)$ as
follows \cite{Parisi}:
\begin{eqnarray}
Z_{\sf inst}(q,\epsilon_1,\epsilon_2)&=&\sum_{k=0}^{\infty}\frac{q^k}{k!}
\left(\frac{\epsilon_1+\epsilon_2}{
\epsilon_1\epsilon_2}\right)^k\nonumber\\
&\times&\int\limits_{\mathbb{R}} \prod_{I=1}^k\frac{d\phi_I}{2\pi i}\int{\cal
D}\rho(\phi)\delta\left(
\rho(\phi)-\sum_{I=1}^k
\frac{\epsilon_1\epsilon_2}{\epsilon_1+\epsilon_2}
\,\delta(\phi-\phi_I)
\right)
\nonumber\\
&
\!\!\!\!\!\!\!\!\!\!\!\!\!\!\!\!\!\!\!\!\!\!\!\!\!\!\!\!\!\!
\!\!\!\!\!\!\!\!\!\!\!\!\!\!\!\!\!\!\!\!\!\!\!\!\!\!\!\!\!\!\!\!
\!\!\!\!\!\!\!\!\!\!\!\!\!\!\!\!\!\!\!\!
\times&
\!\!\!\!\!\!\!\!\!\!\!\!\!\!\!\!\!\!\!\!\!\!
\!\!\!\!\!\!\!\!\!\!\!\!\!\!\!\!\!\!\!\!\!\!\!\!
{\rm e}^{\left[
\fint\limits_{-\infty}^{+\infty}d\phi
d\phi'\left(\frac{\epsilon_1+\epsilon_2}{
\epsilon_1\epsilon_2}\right)^2\rho(\phi)\log(D(\phi-\phi'))\rho(\phi')
+\fint\limits_{-\infty}^{+\infty}d\phi\left(\frac{\epsilon_1+\epsilon_2}{
\epsilon_1\epsilon_2}\right)\rho(\phi)\log(Q(\phi))
\right]}.
\label{were0}
\end{eqnarray}
In the above equation the principal value prescription of the integrals
over the variable $\phi$ is necessary in order to avoid the
singularities of $\log(D(\phi-\phi'))$ when $\phi=\phi'$.
This is the analog in the continuous case of the condition
$I\ne J$ in the double discrete sum appearing in $H_k$.
After introducing the Fourier representation of the Dirac delta
function appearing in the above equation:
\begin{eqnarray}
\delta\left(
\rho(\phi)-\sum_{I=1}^k
\frac{\epsilon_1\epsilon_2}{
  \epsilon_1+\epsilon_2}
\delta(\phi-\phi_I)
\right)
&=&
\nonumber\\
&
&
\!\!\!\!\!\!\!\!\!\!\!\!\!\!\!\!\!\!\!\!\!\!\!\!\!\!\!\!\!\!\!\!\!\!\!
\!\!\!\!\!\!\!\!\!\!\!\!\!\!\!\!\!\!\!\!\!\!\!\!\!\!\!\!\!\!\!\!\!\!\!\!\!
\!\!\!\!\!\!\!\!\!\!\!\!\!\!\!\!\!\!\!\!\!\!\!\!\!\!\!\!\!\!\!\!\!\!\!\!
\int{\cal D}\lambda
\exp\left[
i\int\limits_{-\infty}^{+\infty}d\phi
\lambda(\phi)\left(
\rho(\phi)-\sum_{I=1}^k
\frac{\epsilon_1\epsilon_2}{
  \epsilon_1+\epsilon_2}
\,\delta(\phi-\phi_\sigma)
\right)
\right]\label{ntntn}
\end{eqnarray}
we obtain:
\begin{eqnarray}
Z_{\sf inst}(q,\epsilon_1,\epsilon_2)&=&\sum_{k=0}^{\infty}\frac{q^k}{k!}\int{\cal
D}\rho{\cal D}\lambda \left(
\int\limits_{\mathbb{R}}\frac{d\phi}{2\pi i}\;
\frac{(\epsilon_1+\epsilon_2)}{\epsilon_1\epsilon_2}\,
{\rm e}^{-i \frac{\epsilon_1\epsilon_2}{\epsilon_1+\epsilon_2}
\lambda(\phi)}
\right)^k\nonumber\\
&
\!\!\!\!\!\!\!\!\!\!\!\!\!\!\!\!\!\!\!\!\!\!\!\!\!\!\!\!\!\!\!\!\!\!\!\!
\!\!\!\!\!\!\!\!\!\!\!\!\!\!\!\!\!\!\!\!\!\!\!\!\!\!\!\!\!\!\!\!\!\!\!\!
\!\!\!\!\!\!\!\!\!\!\!\!\!\!\!\!\!\!
\times&
\!\!\!\!\!\!\!\!\!\!\!\!\!\!\!\!\!\!\!\!\!\!\!\!\!\!\!\!\!\!\!\!\!\!\!\!
\!\!\!\!\!\!\!\!\!\!\!\!\!\!
\exp\left[
\fint\limits_{-\infty}^{+\infty}d\phi
d\phi'
\left(\frac{\epsilon_1+\epsilon_2}{\epsilon_1\epsilon_2}\right)^2
\rho(\phi)\log(D(\phi-\phi'))\rho(\phi')\right.\nonumber\\
&\!\!\!\!\!\!\!\!\!\!\!\!\!\!\!\!\!\!\!\!\!\!\!\!\!\!\!\!\!\!\!\!\!\!\!\!
\!\!\!\!\!\!\!\!\!\!\!\!\!\!\!\!\!\!\!\!\!\!\!\!\!\!\!\!\!\!\!\!\!\!\!\!
\!\!\!\!\!\!\!\!\!\!\!\!\!\!\!\!\!\!
+&
\!\!\!\!\!\!\!\!\!\!\!\!\!\!\!\!\!\!\!\!\!\!\!\!\!\!\!\!\!\!\!\!\!\!\!\!
\!\!\!\!\!\!\!\!\!\!\!\!\!\!
\left.
\fint\limits_{-\infty}^{+\infty}d\phi
\left(i\lambda(\phi)\rho(\phi)+
\frac{(\epsilon_1+\epsilon_2)}{\epsilon_1\epsilon_2}
\rho(\phi)
\log(Q(\phi))\right)
\right].
\label{were}
\end{eqnarray}
Let us note that the contour integration over $\mathbb{R}$ is now
restricted to the integral of the quantity ${\rm e}^{-i\lambda(\phi)}$. As
mentioned in \cite{Parisi}, the choice of the contour is not
important when performing integrals over
densities like $\rho(\phi)$ or
$\lambda(\phi)$, so that it is possible to  replace in
Eq.~(\ref{were})
the contour integration with the integration over the whole real line,
i.~e.:
$
\int_\mathbb{R}\;\longrightarrow\fint_{-\infty}^{+\infty}
$.

The summation over the
 index $k$ in Eq.~(\ref{were})
can be easily performed and gives
as a result:
\begin{eqnarray}
&\!\!\!\!\!\!\!\!\!\!
&\!\!\!\!\!\!\!\!\!\!\!\!\!\!\!\!\!\!\!\!
Z_{\sf inst}(q,\epsilon_1,\epsilon_2)=\nonumber\\
&&\!\!\!\!\!\!\!\!\!\!\!\!\!\!\!\!\!\!\!\!\int{\cal
D}\rho{\cal D}\lambda
\exp
\left\{
\fint\limits_{-\infty}^{+\infty}d\phi
d\phi'
\left(\frac{\epsilon_1+\epsilon_2}{\epsilon_1\epsilon_2}\right)^2
\rho(\phi)
\log\left(D(\phi-\phi')\right)
\rho(\phi')\right.\nonumber\\
&\!\!\!\!\!\!\!\!\!\!\!\!\!\!\!\!\!\!\!\!
\!\!\!\!\!\!\!\!\!\!\!\!\!\!\!
+&\!\!\!\!\!\!\!\!\!\!
\!\!\!\!\!\!\!\!\!\!\left.
\fint\limits_{-\infty}^{+\infty}d\phi
\left[i\rho(\phi)\lambda(\phi)+
\left(\frac{\epsilon_1+\epsilon_2}{\epsilon_1\epsilon_2}\right)\left(
\rho(\phi)
\log(Q(\phi))+q\,{\rm e}^{-i
\frac{\epsilon_1\epsilon_2}{\epsilon_1+\epsilon_2}
\lambda(\phi)}\right)\right]
\right\}.\label{were2}
\end{eqnarray}
For future purposes, it will be convenient to perform in Eq.~(\ref{were2})
the following shift of the auxiliary field $\lambda(\phi)$:
\begin{equation}
\lambda(\phi)=\lambda'(\phi)-i\frac{(\epsilon_1+\epsilon_2)}
{\epsilon_1\epsilon_2} \log(q).\label{shift}
\end{equation}
After the above shift, the instanton partition function
$Z_{\sf inst}(q,\epsilon_1,\epsilon_2)$ takes the form:
\begin{eqnarray}
Z_{\sf inst}(q,\epsilon_1,\epsilon_2)&=&\nonumber\\
&&
\!\!\!\!\!\!\!\!\!\!\!\!\!\!\!\!\!\!\!\!\!\!\!\!\!\!\!\!\!\!\!\!\!\!\!\!\!\!\!
\!\!\!\!\!\!\!\!\!\!\!\!\!\!\!\!\!
\int{\cal D}\rho{\cal D} \lambda'
\exp\left\{
\fint\limits_{-\infty}^{+\infty}d\phi
d\phi'
\left(\frac{\epsilon_1+\epsilon_2}{\epsilon_1\epsilon_2}\right)^2
\rho(\phi)
\log\left(D(\phi-\phi')\right)
\rho(\phi')\right.\nonumber\\
&\!\!\!\!\!\!\!\!\!\!\!\!\!\!\!\!\!\!\!\!\!\!\!\!\!\!\!\!\!\!\!\!\!\!\!\!\!\!\!
\!\!\!\!\!\!\!\!\!\!\!\!\!\!\!\!\!
\!\!\!\!\!\!\!\!\!\!\!\!\!\!\!\!\!\!\!\!\!\!\!\!\!\!\!\!\!\!\!\!\!\!\!\!\!\!\!
\!\!\!\!\!\!\!\!\!+&
\!\!\!\!\!\!\!\!\!\!\!\!\!\!\!\!\!\!\!\!\!\!\!\!\!\!\!\!\!\!\!\!\!\!\!\!\!
\!\!\!\!\!\!\!\!\!\!\!\!\!\!
\!\!\!\!\!\!
\left.
\fint\limits_{-\infty}^{+\infty}d\phi
\left[i\rho(\phi)\lambda'(\phi)+
\left({\textstyle\frac{\epsilon_1+\epsilon_2}{\epsilon_1\epsilon_2}}\right)\left(
\rho(\phi)
\log(qQ(\phi))+{\rm e}^{-i
\frac{\epsilon_1\epsilon_2}{\epsilon_1+\epsilon_2}
\lambda'(\phi)}\right)\right]\nonumber
\right\}.\nonumber\\
\label{were2bis}
\end{eqnarray}
This is the desired path integral expression of the instanton
partition function.

Let us note that we could have arrived to Eq.~(\ref{were2bis})
starting from Eq.~(\ref{were0}) and performing there
the substitution:
\begin{equation}
q^k= \exp
\left[\frac{\epsilon_1+\epsilon_2}{
\epsilon_1\epsilon_2}\sum_{I=1}^k\frac{\epsilon_1\epsilon_2}{
\epsilon_1+\epsilon_2}\log(q)\right].
\end{equation}
 This replacement obviously does
not change 
Eq.~(\ref{were0}) because it is easy to prove
the following sequence of identities:
\begin{equation}
q^k=\textrm{e}^{k\log(q)}=\textrm{e}^{\sum_{I=1}^k\log(q)}=\textrm{e}^
{\frac{\epsilon_1+\epsilon_2}{
\epsilon_1\epsilon_2}\sum_{I=1}^k\frac{\epsilon_1\epsilon_2}{
\epsilon_1+\epsilon_2}\log(q)}.
\end{equation}
Using the condition (\ref{eigdensdef}) that is imposed
by the Dirac delta function appearing in Eq.~(\ref{were0}),
we may finally write:
\begin{equation}
q^k=
\exp\left(\frac{\epsilon_1+\epsilon_2}{\epsilon_1\epsilon_2}
\fint\limits_{-\infty}^{+\infty}d\phi\rho(\phi)\log(q)\right).\label{sdsdssd}
\end{equation}
In this way Eq.~(\ref{were0}) becomes:
\begin{eqnarray}
Z_{\sf inst}(q,\epsilon_1,\epsilon_2)&=&\sum_{k=0}^{\infty}\frac{1}{k!}
\left(\frac{\epsilon_1+\epsilon_2}{
\epsilon_1\epsilon_2}\right)^k\nonumber\\
&\times&\int\limits_{\mathbb{R}} \prod_{I=1}^k\frac{d\phi_I}{2\pi i}\int{\cal
D}\rho(\phi)\delta\left(
\rho(\phi)-\sum_{I=1}^k
\frac{\epsilon_1\epsilon_2}{\epsilon_1+\epsilon_2}
\,\delta(\phi-\phi_I)
\right)
\nonumber\\
&
\!\!\!\!\!\!\!\!\!\!\!\!\!\!\!\!\!\!\!\!\!\!\!\!\!\!\!\!\!\!
\!\!\!\!\!\!\!\!\!\!\!\!\!\!\!\!\!\!\!\!\!\!\!\!\!\!\!\!\!\!\!\!
\!\!\!\!\!\!\!\!\!\!\!\!\!\!\!\!\!\!\!\!
\times&
\!\!\!\!\!\!\!\!\!\!\!\!\!\!\!\!\!\!\!\!\!\!
\!\!\!\!\!\!\!\!\!\!\!\!\!\!\!\!\!\!\!\!\!\!\!\!
{\rm e}^{\left[
\fint\limits_{-\infty}^{+\infty}d\phi
d\phi'\left(\frac{\epsilon_1+\epsilon_2}{
\epsilon_1\epsilon_2}\right)^2\rho(\phi)\log(D(\phi-\phi'))\rho(\phi')
+\fint\limits_{-\infty}^{+\infty}d\phi\left(\frac{\epsilon_1+\epsilon_2}{
\epsilon_1\epsilon_2}\right)\rho(\phi)\log(qQ(\phi))
\right]}.
\label{were01}
\end{eqnarray}
After exploiting the Fourier representation of the Dirac delta function
given in Eq.~(\ref{ntntn}),
it is possible to obtain Eq.~(\ref{were2bis}) directly from
(\ref{were01}) without performing the shift (\ref{shift}).
As a matter of fact,
the right hand side of Eq.~(\ref{sdsdssd}) provides exactly the term
proportional to
$\log(q)$ that is present in (\ref{were01}) but is apparently missing
in the action of
Eq.~(\ref{were2}) and has been recovered only after the shift
(\ref{shift}).

In order to pass to the limit $\epsilon_2\to0$ in
Eq.~(\ref{were2bis}), the following two
formulas will be useful:
\begin{eqnarray}
&&\fint\limits_{-\infty}^{+\infty} d\phi
d\phi'\rho(\phi)\log(D(\phi-\phi')) \rho(\phi')=\nonumber\\
&&\fint\limits_{-\infty}^{+\infty} d\phi\,\fint\limits_{-\infty}^{\phi}
d\phi'\rho(\phi)
\left[
\log(D(\phi-\phi'))-\log(D(\phi'-\phi))
\right]
\rho(\phi')\label{aac}
\end{eqnarray}
and
\begin{equation}
\lim_{\epsilon_2\to 0}\left[\frac{
\log(D(\phi-\phi'))+\log(D(\phi'-\phi))
}{\epsilon_2}\right]=\frac{2\epsilon_1}{\epsilon_1^2-(\phi-\phi')^2}.
\label{aad}
\end{equation}
With the help of equations (\ref{aac}) and (\ref{aad})
 it is easy to prove that, when
$\epsilon_2\sim 0$, the expression of $Z_{\sf
  inst}(q,\epsilon_1,\epsilon_2)$
given in Eq.~(\ref{were2bis}) may be
approximated up to an irrelevant constant as shown below:
\begin{eqnarray}
&&Z_{\sf inst}(q,\epsilon_1,\epsilon_2)\sim
\nonumber
\\
&&\hspace{10pt}
\int{\cal D}\rho(\phi)\,\exp\left\lbrace
\frac 1{\epsilon_2}
\left[\frac 12\fint\limits_{-\infty}^{+\infty}d\phi
\fint\limits_{-\infty}^{+\infty}d\phi'\rho(\phi)G(\phi-\phi')
\rho(\phi')\right.\right.\nonumber\\
&+&\left.\left.\fint\limits_{-\infty}^{+\infty}d\phi\rho(\phi)
\log(qQ_0(\phi))
\right]\right\rbrace,\label{were3}
\end{eqnarray}
where
\begin{equation}
G(\phi-\phi')=\frac{2\epsilon_1}{\epsilon_1^2-(\phi-\phi')^2}.
\label{aae}
\end{equation}
Let us notice that the dependence on the auxiliary field
$\lambda(\phi)$ disappeared after the limit $\epsilon_2\to 0$.
The above equation is in agreement with the saddle--point
approximation that have been derived in the literature, see for
example Ref.~\cite{Fucitoiinni}.
In order to derive Eq.~(\ref{were3}) we have used the fact that:
\begin{equation}
\fint\limits_{-\infty}^{+\infty}d\phi
\fint\limits_{-\infty}^{\phi}d\phi'\rho(\phi)G(\phi-\phi')
\rho(\phi')=
\frac 12\fint\limits_{-\infty}^{+\infty}d\phi\,
\fint\limits_{-\infty}^{+\infty}d\phi'\rho(\phi)G(\phi-\phi')
\rho(\phi').
\end{equation}
\section{Conclusions}
In this work a path integral formulation of the full Nekrasov
instanton partition function has been derived using the methods of
matrix models, see Eq.~(\ref{were2bis}).
As a byproduct, it has been
shown in a rigorous and simple way that
in the semi-classical limit, valid when $\epsilon_2$ is very small, the
leading order contribution in the partition function is provided by
Eq.~(\ref{were3}), in agreement with the argument of
\cite{NekrasovOkounkov} and statistical mechanics.

In the future it is planned to study the path integral appearing in
Eq.~(\ref{were2bis}) using the method of Ref.~\cite{FFgenpot} that
allows to perform analytic calculations in the case of theories with
nonpolynomial and complex potentials.
\section{Acknowledgments}
This research has been supported in part by the Polish National
Science Centre under Grant No. N202 326240.

\end{document}